\documentclass[twocolumn,aps,prl,groupedaddress,showpacs, superscriptaddress]{revtex4-2} 

\usepackage{amsmath}
\usepackage{amssymb}
\usepackage{gensymb}
\usepackage{graphicx,bm,units,yfonts}
\usepackage[table]{xcolor}
\usepackage{hyperref}
\usepackage{movie15}
\usepackage{color}

\newcommand{\Kk}{{\mathcal K}}

\newcommand{\Ll}{{\mathcal L}}

\begin{document}

\title{Topological D-Class Physics with Passive Acoustic Elements}

\author{Wenting Cheng}
\affiliation{Department of Physics, New Jersey Institute of Technology, Newark, NJ, USA}

\author{Emil Prodan}
\email{prodan@yu.edu}
\affiliation{Department of Physics, Yeshiva University, New York, NY, USA}

\author{Camelia Prodan}
\email{cprodan@njit.edu}
\affiliation{Department of Physics, New Jersey Institute of Technology, Newark, NJ, USA}

\begin{abstract} In this work, we supply an engineering solution that paves the way to the implementation of the full classification table of topological insulators with passive acoustic crystals. As an example, we design an acoustic crystal displaying the full range of characteristics of a topological insulator from class D, such as non-trivial Chern numbers and an exact particle-hole symmetry for both bulk and edge dynamics. The key is a particular geometry of the acoustic resonators that facilitates a large number of independent and equal-strength couplings, eight to be precise, such that the mode-coupling theory remains accurate. The latter enables us to implement with high fidelity the de-complexified tight-binding model generating the topological phases from class D in 2-dimensions, previously announced in \cite{BarlasPRB2018}. The successful realization of a topological acoustic crystal from class D paves the way to the observation of the Majorana-like modes stabilized by $\pi$-fluxes.
\end{abstract}

\maketitle

The discovery of topological insulators \cite{HaldanePRL1988,KanePRL2005A,KanePRL2005B,BernevigScience2006,
KonigScience2007,MoorePRB2007,FuPRB2007A,FuPRB2007B} has revolutionized the design principles of solid state materials and devices. One of the most important achievements in the field was the completion of the classification table that enumerates the strong topological insulators \cite{SchnyderPRB2008,QiPRB2008,Kitaev2009,RyuNJP2010}. Their different phases are separated by true phase boundaries where a localization-delocalization phase transition occurs and they also display bulk-boundary correspondences that are immune to strong disorder \cite{ProdanJPA2011}. The entries from the table sample the ten fundamental Altland-Zirnbauer universality classes \cite{AltlandPRB1997} of mesoscopic systems, hence the topological phase transitions have unique signatures in each class. Furthermore, each of the ten classes supports unique physical phenomena and, for these reasons, it is important to have reliable materials which sample the entire classification table. In particular, topological insulators from class D are interesting because they support majorana excitations, which can be stabilized and braided by $\pi$-flux Dirac solenoids \cite{LiuAOP2020}.

Around the time when the classification table was perfected, similar topological phases were predicted for classical degrees of freedom, both electromagnetic \cite{HaldanePRL2008} and mechanical \cite{ProdanPRL2009}. The second work observed that quadratic terms like $q\cdot p$ in a Lagrangean break the time-reversal symmetry of a system and can drive it into a topological Chern insulating phase. Such terms require active elements and, indeed, the first laboratory realization of a mechanical Chern insulator was based on active components that break the time-reversal symmetry via such mechanism \cite{NashPANS2015}. An interesting alternative for producing mechanical topological phases came soon after in \cite{SusstrunkScience2015}, where just passive but well thought couplings together with doubling the degrees of freedom creates the topological effect. A design based exclusively on real couplings has the tremendous advantage that no fine-tuning and enforcement of any phase delay is required. In acoustics, such designs will require just in- and out-of phase couplings of the resonators, which can be achieve with short acoustic bridges. This will free the designs from the use of chiral tubes \cite{Note0}, which deliver the needed functionality only at one specific frequency and, as such, make the implementation of the particle-hole symmetry impossible because the latter manifests over an entire range of frequencies. Therefore, a challenge was put forward to implement the whole classification table of topological insulators exclusively with passive elements and real couplings. In two and three dimensions, to authors' knowledge, none of the topological phases from the table have been implemented with classical degrees of freedom, except for those from A and AII classes.

The challenge was met in \cite{BarlasPRB2018} with a simple theoretical solution. The model Hamiltonians from the classification table \cite{RyuNJP2010} contain complex coefficients, hence they need to be de-complexified without altering the algebraic operations. The latter is need in order to preserve both the bulk and boundary spectra. This was achieved in \cite{BarlasPRB2018} by replacing $\sqrt{-1}$ by the real matrix $\big (\, ^0_1 \ ^1_0 \big )$ and this substitution maps any complex $H=\sum \hat w_{\bm x,\bm x'} \otimes |\bm x\rangle \langle \bm x' |$ defined over a lattice $\Ll$ to a real Hamiltonian
\begin{equation}\label{Eq:RealH}
\rho(H) = \sum_{\bm x,\bm x' \in \Ll} \begin{pmatrix} {\rm Re}[\hat w_{\bm x,\bm x'}] & {\rm Im}[\hat w_{\bm x,\bm x'}] \\ -{\rm Im}[\hat w_{\bm x,\bm x'}] & {\rm Re}[\hat w_{\bm x,\bm x'}] \end{pmatrix} \otimes |\bm x \rangle \langle \bm x' |,
\end{equation}
at the expense of doubling the degrees of freedom.

\begin{figure*}[t!]
\includegraphics[width=\linewidth]{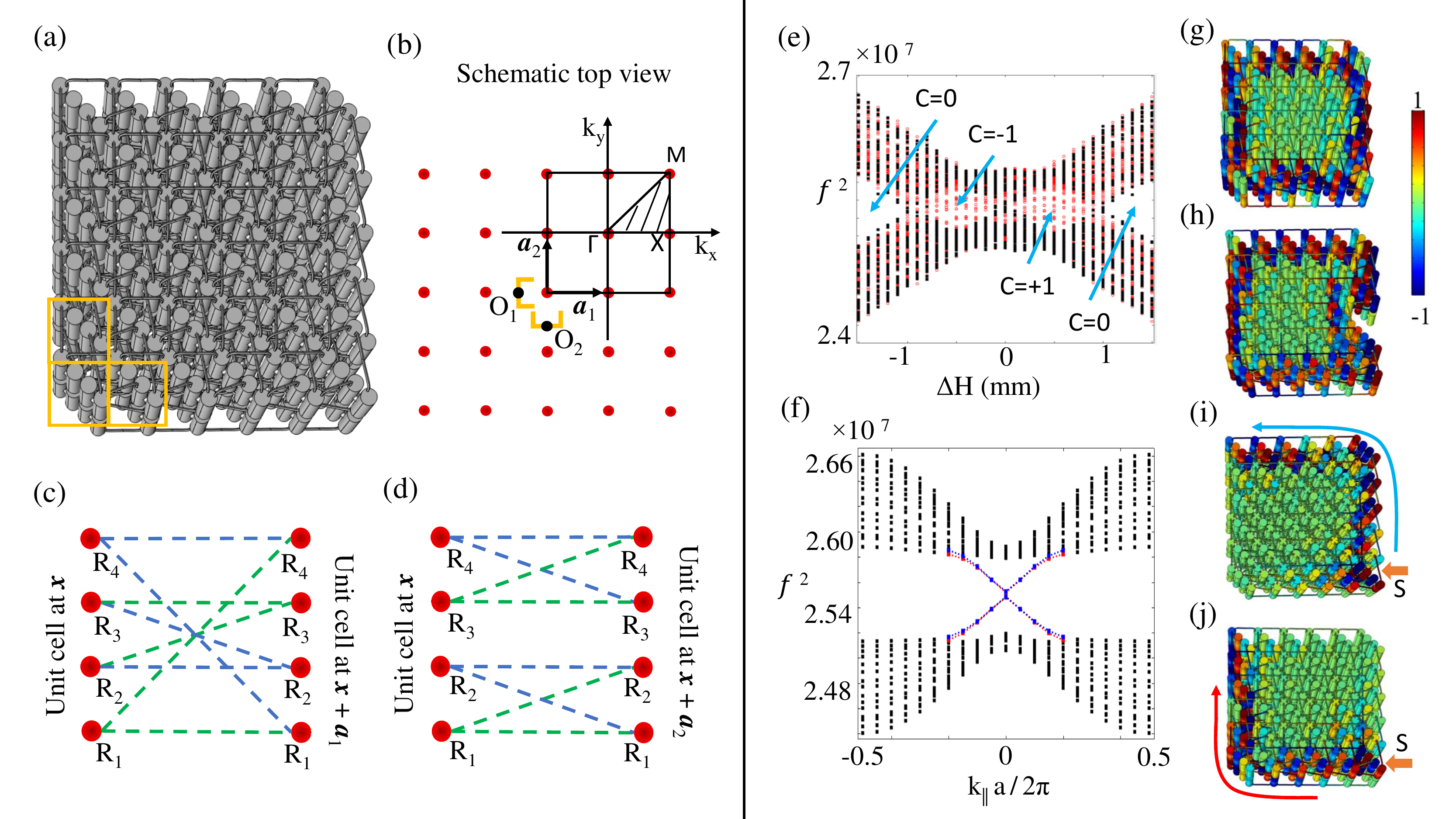}
\caption{\small {\bf Acoustic crystal implementing the Hamiltonian from Eq.~\eqref{Eq:Model1} with passive acoustic elements:}  (a) The actual domain of the acoustic equation used in the COMSOL simulation of a system composed of 6 $\times$ 6 unit cells. The yellow boxes indicate three unit cells $\bm x$, $\bm x+\bm a_1$ and $\bm x+\bm a_2$, to be referred to in panels (c) and (d).  (b-d) Schematic representation of the lattice: (b) Top view of the square lattice with two observers $O_1$ and $O_2$ looking along the $\bm a_1$ and $\bm a_2$ directions, respectively. Each red dot represents four resonators, which in this schematic have been stack on top of each other. (c,d) The couplings implementing $\rho(H)$, as seen by the observers $O_1$ and $O_2$, respectively. The blue dashed lines indicate the positive couplings and the green dashed lines indicate the negative couplings, all of equal strengths. Each red dot represents one resonator. (e) COMSOL simulation of the resonant spectrum of the crystal in a 8-cells wide ribbon geometry with periodic (black markers) and hard (red markers) boundary conditions in the transversal direction as function of $\Delta H = H_1 - H_2$ defined in Fig.~3 and representing the mass parameter from Eq.~\eqref{Eq:Model1}. The simulation is in perfect agreement with the known phase diagram of model~\eqref{Eq:Model1}. (f) Resonant spectrum corresponding to $\Delta H =0.5$mm, rendered as function of the quasi-momentum parallel to the ribbon. Bulk spectrum (black dots) and two doubly-degenerate chiral edge bands (marked in red and blue) are visible, displaying a high degree of mirror symmetry relative to the mid-gap. (g,h) COMSOL simulations of the mode profiles corresponding to a frequency inside the bulk gap ($\approx 5040$Hz) and finite geometries with clean edges and with a defect, respectively. (i) Sound excitation with a source (see arrow) at 5035.1Hz and with sound-absorbing boundary conditions on the left and lower edges. The latter filter out one topological channels, leaving only the uni-directional anti-clockwise sound propagation. (j) Same as (i), only with sound-absorbing boundary conditions imposed to the right and upper edges, prompting clockwise unidirectional wave propagation.} 
\label{Fig:ChernInsulator}
\end{figure*}

Any such Hamiltonian has a built-in $U(1)$-symmetry, since $\rho(H)$ automatically commutes with $
U = \begin{pmatrix} ^0_{-\mathbb I}  & ^{\mathbb I}_0 \end{pmatrix} \otimes I$, and also has a fermionic time reversal symmetry $\Sigma = U \mathcal K = \mathcal K U$, $\Sigma^2 = - 1$ ($\mathcal K$ = complex conjugation). The matrix $U$ has spectrum at $\pm 1$ and we let $\Pi_\pm$ be the projectors onto the corresponding spectral subspaces. Now, assume that the original complex Hamiltonian displays one of the symmetries from the classification table, $\Theta H \Theta^{-1} =\epsilon H$, $\epsilon=\pm 1$, with $\Theta$ anti-unitary and $\Theta^2 = \gamma$, $\gamma= \pm 1$. Such $\Theta$ can always be written as $\Kk W = W \Kk$ with $W$ unitary and, if $\rho_\pm(H) = \Pi_\pm \rho(H) \Pi_\pm$, the following remarkable identity holds \cite{BarlasPRB2018}
  \begin{equation}
	\label{Eq:mappedsymmetry}
\epsilon \, \rho_\pm(H)  = \big (\Kk \rho(W) J\big ) \rho_\pm (H)  \big ( \Kk \rho(W) J \big )^{-1},
 \end{equation}
where $
J = \begin{pmatrix} ^0_{\mathbb I}  & ^{\mathbb I}_0 \end{pmatrix} \otimes I$. In other words, the symmetry projected real Hamiltonian obeys exactly the same symmetry as the complex operator, which is implemented by
\begin{equation}\label{Eq:Signature}
\tilde \Theta = \Kk \, \rho(W) J, \quad \tilde \Theta^2 =\gamma.
\end{equation}
Note that the original symmetries $U$ and $\Sigma$ have little to do with $\tilde \Theta$. Ref.~\cite{BarlasPRB2018} checked the above statement as well as the bulk-boundary correspondences for all model Hamiltonians from the table up to dimension three. The dynamics of these Hamiltonians with real coefficients can be projected on the symmetry sectors $\Pi_\pm$ by choosing appropriate initial conditions or using specialized sources \cite{BarlasPRB2018}. More practical ways to project onto the symmetry sectors will be discussed below.

The above principles have been around for several years, but experimental implementations resisted to show up. Although optimal, the map~\eqref{Eq:RealH} leads to relatively complex crystal structures. For orientation, the Kane-Mele model with Rashba coupling \cite{KanePRL2005A,KanePRL2005B}, hence the true topological insulator from AII class, requires eight resonators per repeating cell \cite{BarlasPRB2018} (please see note~\cite{Footnote1}) and eight connections per resonator. In this work, we make the first important step towards the experimental implementation of the {\it full} classification table with passive components, by demonstrating a topological  acoustic crystals from class D (see Fig.~1), built on the principles described above. There are two main challenges here: 1) opening a topological gap that carries a non-trivial Chern number and 2) enforcing the particle-hole symmetry associated with the class D, for both bulk and boundaries. Our work demonstrates that full control over both aspects can be achieved. Let us stress that the passage from class A to class D is significant because the particle-hole symmetry needs to be enforced over a significant range of frequencies. The engineering challenge is to ensure that mode-coupling theory \cite{Footnote2}, which we use to connect with the theoretical tight-binding models, is accurate. This is difficult when large number of connections are needed.

If $\sigma_i$ denote Pauli's matrices and $S_j|\bm x\rangle = |\bm x + \bm a_j\rangle$ represent the shift operators,  with $\bm a_j$, $j=1,2$, being the generators of the lattice (Fig.~1(b)), the model generating the class D in 2-dimensions reads \cite{RyuNJP2010}
\begin{equation}\label{Eq:Model1}
H = \tfrac{1}{\imath} \sum_{j=1,2} \sigma_j \otimes (S_j - S_j^\dagger) + \sigma_3 \otimes \big [ m + \sum_{j=1,2}(S_j + S_j^\dagger) \big ].
\end{equation}
The particle-hole symmetry is implemented by the anti-unitary operation $\Theta_{\rm PH} = (\sigma_1 \otimes I) \Kk$ and its phase diagram consists of topological phases with Chern number $C=-1$ for $m\in (-4,0)$ and $C=+1$ for $m\in (0,4)$, as well as of a trivial phase $C=0$ for $m \notin [-4,4]$.

\begin{figure}[t!]
\includegraphics[width=\linewidth]{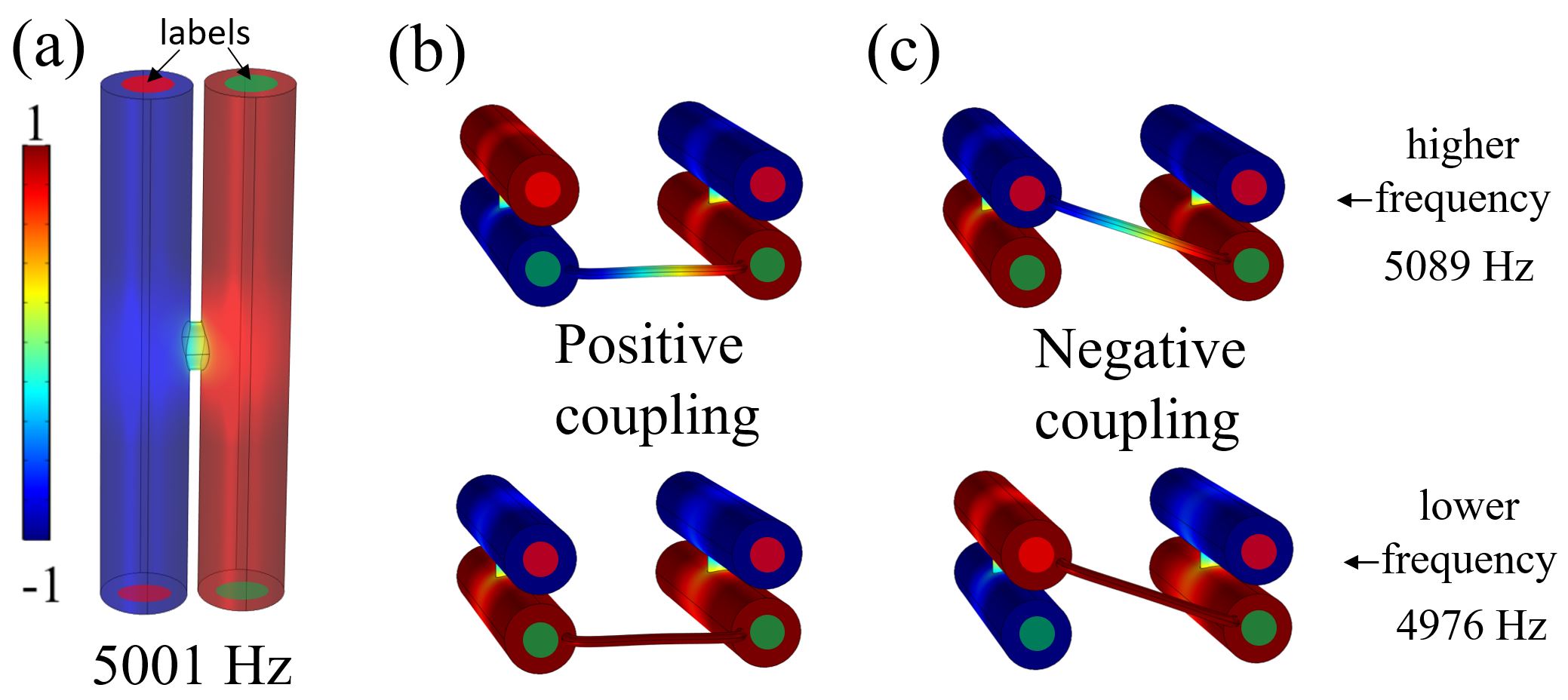}
\caption{\small {\bf H-shaped acoustic resonators and their positive and negative couplings} (a) First resonant mode has frequency $f \approx 5000$Hz and consists of out-of-phase pressure field $|p\rangle$ that is, to a high degree, uniform along the arms of the H-shape. The red and green dots will be used for reference, to communicate the connections in the more schematic representations in Fig.~\ref{Fig:unit cell}. (b) An acoustic bridge between a pair of resonators, which fits the mode-coupling Hamiltonian $f^2|p_1\rangle +f^2|p_2\rangle + \beta (|p_1\rangle \langle p_2 | +|p_2\rangle \langle p_1|)$ with a positive coupling coefficient $\beta$. (c) An acoustic bridge leading to a negative $\beta$ but identical coupling strength $|\beta|$.} 
\label{Fig:Connections}
\end{figure}

\begin{figure}[t!]
\includegraphics[width=\linewidth]{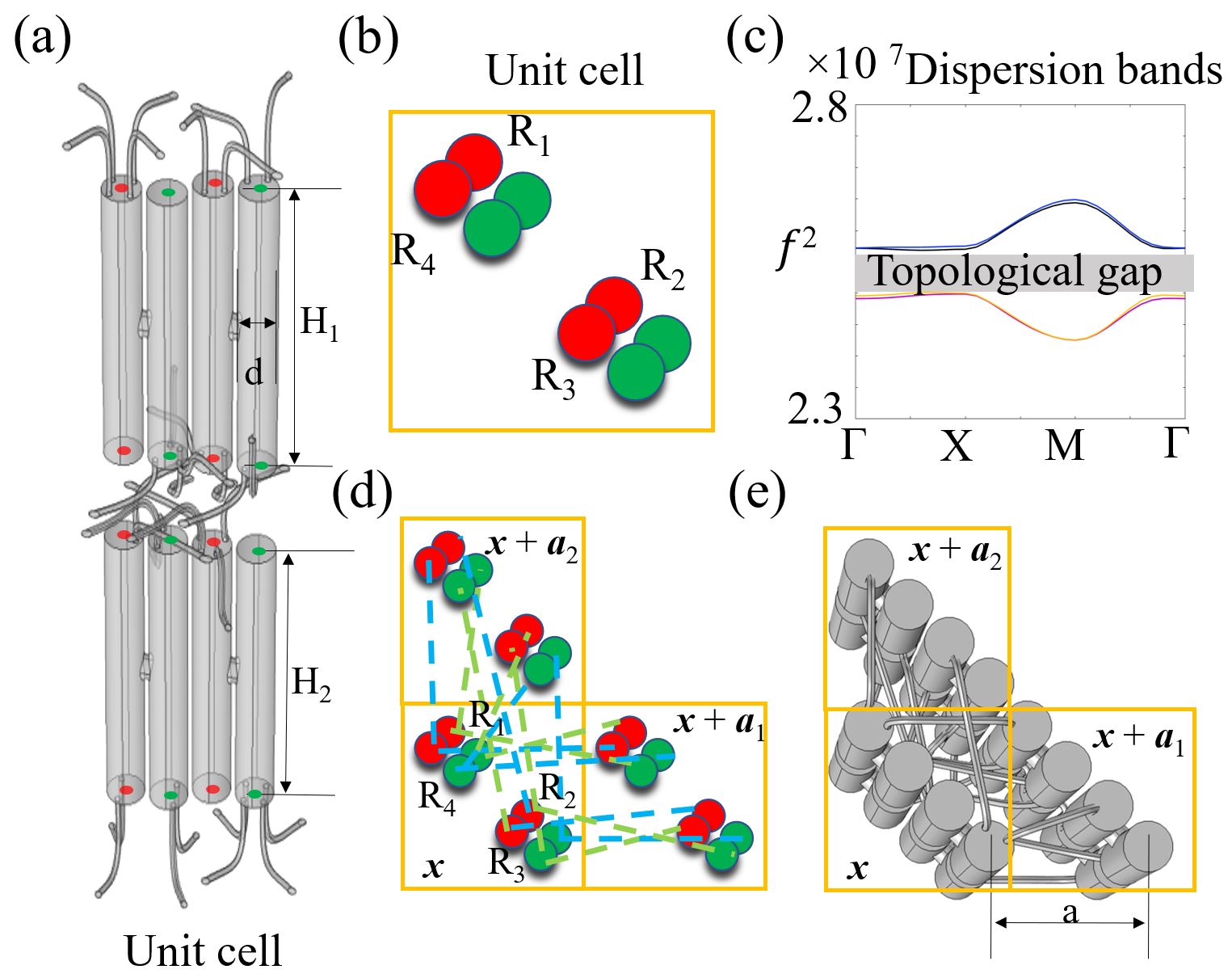}
\caption{\small {\bf Unit cell design of the topological acoustic crystal} (a) The domain of the acoustic wave equation used in the COMSOL simulations, corresponding to the unit cell. It consists of four H-shape resonators placed in two rows and two columns. (b) Schematic of the unit cell showing only the labels (red/green dots) of the arms. (c) COMSOL simulation of the bulk band structure along the high symmetry lines of the first Brillouin zone of a square lattice, revealing a topological bandgap and almost perfect doubly degenerate bands that are mirror symmetric relative to the mid-gap. (d,e) Schematic and realistic representations of the theoretical couplings specified in Fig.~\ref{Fig:ChernInsulator}(c,d). Note that some of the couplings are at the top and and some at the bottom of the H-shaped resonators.} 
\label{Fig:unit cell}
\end{figure}

The mapping~\eqref{Eq:RealH} leads to a crystal with four resonators in the primitive cell and the connections shown in Figs.~\ref{Fig:ChernInsulator}(b-d). Each resonator needs a total of eight connections of equal strength with its neighboring resonators. Implementing these connections was one of the main design challenges. A search among various options led us to the H-shaped resonator shown in Fig.~\ref{Fig:Connections}, which is quite unique. Its first resonating mode, shown in Fig.~\ref{Fig:Connections}(a), displays a uniform pressure field along the long arms. Positive and negative couplings between these resonant modes carried by two resonators can be established by the short bridges shown in Figs.~\ref{Fig:Connections}(b,c). The main advantage of the H-shaped resonators and the use of the first resonant mode is that the bridges can be attached to four ends, hence reducing the load per end, and can be moved between the arms of the H-shape sharing the same color, without affecting the strength and sign of the coupling. This feature ultimately enabled us to implement the eight needed connections of equal strengths and desired signs. The details of the primitive cells and of the specific connections are supplied in Fig.~\ref{Fig:unit cell}. The mass term in Eq.~\eqref{Eq:Model1} is implemented by a mismatch $\Delta H = H_1 -H_2$ between the heights of resonators located inside the same primitive cell (see Fig.~\ref{Fig:unit cell}(a)). The geometric parameters are as follows: The lattice constant of the square lattice in xy-plane is $a=12.73$mm. The diameter of the long arms of the H-shaped resonator is $d=4$mm and, in all experiments except for those reported in Fig.~\ref{Fig:ChernInsulator}(e), their heights were chosen as $H_1=22$mm for R$_1$, R$_3$ and $H_2=21.5$mm for R$_2$, R$_4$. All coupling bridges have a length $20$mm and diameter $0.6$mm.

We now demonstrate that the acoustic crystal indeed delivers the expected physics from the topological D class.  Fig.~\ref{Fig:ChernInsulator}(e) reports COMSOL simulations of the resonant spectrum of the crystal in a ribbon geometry as function of the mass parameter $\Delta H$. When periodic boundary conditions are imposed in the transversal direction, a clean gap opens and closes in the bulk spectrum in full agreement with the phase diagram of the model mentioned earlier. Furthermore, when hard boundary conditions are used, edge spectrum shown with red markers in Fig.~\ref{Fig:ChernInsulator}(e) fills entirely the bulk gaps in the topological sides of the phase diagram, while being completely absent in the non-topological side. Fig.~\ref{Fig:ChernInsulator}(f) reports the resonant spectrum of the crystal in the same ribbon geometry and for the specific value $\Delta H=0.5$mm, which is in the middle of a topological phase. This time, the spectrum is rendered as function of the quasi-momentum parallel to the ribbon. The connecting bridges intersecting the transversal boundaries of the ribbon have been removed entirely as, for example, visible in panels (g-h) of same figure. The expected chiral edge bands of opposite slopes and localized on opposite boundaries of the ribbon are  indeed present. As clearly visible in Fig.~\ref{Fig:ChernInsulator}(f), these chiral bands are doubly degenerate, which is the proof for the enforcement of the symmetry $U$ mentioned earlier. Another remarkable feature of the reported spectrum is the almost perfect mirror symmetry with respect to the mid-gap line, which demonstrates the expected particle-hope symmetry for both the bulk and edge states. Thus, the COMSOL simulations reproduce all the expected features derived from $\rho(H)$.

Fig.~\ref{Fig:ChernInsulator}(g) illustrates the spatial profile of an eigen-mode with eigen-frequency inside the topological bulk gap, for a geometry with clean edges. As expected, the mode is localized at the edge of the crystal and surrounds entirely the crystal, hence supplying a wave channels for sound propagation. Fig.~\ref{Fig:ChernInsulator}(h) demonstrates that this channel is robust against the imperfections of the boundaries, as expected from a strong topological crystal.

The edge channels reported in Figs.~\ref{Fig:ChernInsulator}(g,h) are doubly degenerate and they support clockwise and anti-clockwise propagating modes, which cannot back-scatter due to the $U$ symmetry mentioned earlier. A generic source of sound will excite both modes, hence the sound propagation will not be uni-directional. This brings us to the issue of projecting onto one of the symmetry sectors $\Pi_\pm$. As we already mentioned, this can be done by using specialized sources of sound that excite only modes with desired symmetry. While the design of these sources is known \cite{BarlasPRB2018}, they can be complicated and require fine tuning. In Fig.~\ref{Fig:ChernInsulator}(i,j), we demonstrate a simple principle to filter the clockwise or the anti-clockwise propagating modes by using a generic source of sound and a patch of absorbing boundary placed at one side or the other of the source. This together with the ability of the crystal to support pairs of topological edge channels can prove useful in various applications where on-demand clockwise or anti-clockwise signal propagation is needed.

Looking forward into the near future, we plan to further improve the design and to ultimately demonstrate the concept with a real laboratory model. This will then enable the exploration and confirmation of the unique characteristics of class D topological systems. For example, topological phase transitions revealed in the phase diagram from Fig.~\ref{Fig:ChernInsulator}(e) can be investigated in the presence of disorder and the expected localization-delocalization phase transition can be finally confirmed experimentally. Furthermore, the critical behavior of the physical observables should be in line with the Wigner surmize at $\beta =2$ \cite{AltlandPRB1997}, hence quite distinct from anything reported experimentally so far. We point out the relatively recent success \cite{MeierScience2018}, where the critical behavior for class AIII in dimension one was finally observed with atomic chains.

Another interesting direction enabled by our successful design of a crystal from class D is the observation of majorana-like modes stabilized by $\pi$-fluxes. The explicit tight-binding models were already supplied in \cite{LiuAOP2020} and, by applying the map $\rho$ from Eq.~\eqref{Eq:RealH}, we can generate explicit acoustic crystals that simulate the majorana states. As demonstrated in \cite{LiuAOP2020}, the braiding and fusion of $\pi$-fluxes is non-trivial and worth exploring even at the one-particle level covered by the linear wave equation regime employed here. We mention that, in one dimension, the work on majorana-like modes implemented with classical degrees of freedom is vigorously underway \cite{ChenAM2019,BarlasPRL2020,GuoPRR2021,QianArxiv2022,AlleinArxiv2022}, with the recent works \cite{QianArxiv2022,AlleinArxiv2022} demonstrating experimental control at a level that makes the classical systems feasible for information storage and processing. In 2 dimensions, implementing the $\pi$-fluxes will require non-uniform perturbations of the couplings and this will require a major effort.

Looking further into the future, we are confident that the H-shaped resonators can be successfully employed to assemble acoustic crystals that simulate the remaining classes in the classification table. As we already mentioned, some of the classes will require eight resonators per unit cell \cite{BarlasPRB2018}, but each resonator is coupled with its neighbors in a fashion very similar to what was already implemented in this work, specifically, eight equal-strength connections per resonator. The H-shape can be iterated to generate geometries that expose more legs where robust and effective couplings can be established. This will also answer the more general current challenge in meta-materials science of implementing complex unit cells and connectivities.

\acknowledgements All authors acknowledge support from the National Science Foundation through the grant CMMI-2131760. E. P. acknowledges additional support from the National Science Foundation through the grant DMR-1823800.


\begin{thebibliography}{1}
\bibitem{Footnote2} Mode-coupling theory assumes that the pressure field of a collective resonant acoustic mode over a lattice of coupled resonators is identical with the one of a decoupled resonator when the pressure field is probed inside that resonator. This theory is accurate when the connections have a small footprint and they do not influence each other.
\end{thebibliography}
\end{document}